\documentclass{iopconfser}

\usepackage{amssymb,amsfonts,physics,braket,slashed,amsmath,amsthm,amsbsy,epsfig,color,graphicx,times}
\usepackage[ansinew]{inputenc}
\usepackage[english]{babel}
\usepackage{color}
\usepackage{braket}
\usepackage{tabularx}
\usepackage{array}

\begin{document}

\title{Exploring $X_{17}$ and dark charges in the context of Standard Model tensions}

\author{Antonio Capolupo$^{1}$, Aniello Quaranta$^{2}$ and Raoul Serao$^{1}$}

\affil{$^1$Dipartimento di Fisica ``E.R. Caianiello'' and INFN gruppo collegato di Salerno, Universit\`{a} degli Studi di Salerno, I-84081
Fisciano (SA), Italy}

\vspace{4mm}

\affil{$^2$School of Science and Technology, University of Camerino,
Via Madonna delle Carceri, Camerino, 62032, Italy}

\email{capolupo@sa.infn.it, aniello.quaranta@unicam.it, rserao@unisa.it}

\begin{abstract} We report on recent results \cite{RS} according to which the hypothetical $X_{17}$ boson could affect the muon $g-2$ anomaly and the Lamb shift. Moreover by considering the kinetic mixing between this new boson and the $U(1)_Y$ we establish possible contributions of the $X_{17}$ to the $W$ mass.
\end{abstract}

\section{Introduction}

In recent decades, the Standard Model has received increasing experimental confirmation of its validity. However, many open questions in physics still require explanations that go beyond the Standard Model. These include the origin and nature of dark matter and dark energy, particle mixing and oscillations, and the matter-antimatter asymmetry \cite{BSM1,BSM2,BSM3,BSM4,BSM5,BSM6, BSM7, BSM8, BSM9, BSM10, BSM11, BSM12, BSM13, BSM14, BSM15, BSM16, BSM17, BSM18, BSM19, BSM20, BSM21, BSM22, BSM23, BSM24}.

Additionally, there is a whole class of "tensions" between the Standard Model's predictions and experimental results, which could provide clues on how to move beyond it. One notable example is the muon anomalous magnetic moment  \cite{Marciano2016, Cazzaniga2021, NL}.
Denoting the muon anomalous magnetic moment $a_\mu=\frac{g_\mu-2}{2}$, we analyze the tension between the Standard Model prediction $a_\mu^{SM}$ and the experimental value $a_\mu^{EXP}$.
The Muon g-2 collaboration's latest experimental results \cite{Abi2021} have confirmed a discrepancy for $a_\mu$ with respect to the Standard Model value \cite{Patrignani2016,Aoyama2020}. The combined data from the Brookhaven \cite{Bennett2004} and Fermilab Muon g-2 \cite{Abi2021} experiments indicate a $4.2 \sigma$ discrepancy: $\Delta a_{\mu} = a_{\mu, EXP} - a_{\mu, SM} = (251 \pm 59) \times 10^{-11}$.
A similar analysis could in principle be performed also for the electron anomalous magnetic moment even if the experimental value $a^{EXP}_e=1159652180.73\times 10^{-12}$ is precise to $0.24$ parts-per-billion, leading to $\Delta a_e = a_{e, EXP} - a_{e, SM} = (4.8  \pm 3.0) \times 10^{-13}$\cite{Hannake,Nature2020}.

Another phenomenon that probably needs an explanation beyond the standard model is the so- called Lamb shift for muonic atoms. 
In particular, the Lamb shift between $2S$ and $2P$ levels for muonic hydrogen and muonic deuterium differs from the Standard Model expected values by: $\delta E_\mu^H=(-0.363,-0.251)\ \mathrm{meV}$ \cite{muon1, muon2} and  $\delta E_\mu^D=(-0.475,-0.337) \ \mathrm{meV}$ \cite{muon3, muon4}. 
Another intriguing hint for new physics is related to the mass of the $W$ boson.
Recently, the CMS collaboration reported an accurate measurement of the $W$ mass, which is $80360 \pm 9.9 \ \mathrm{MeV}$ \cite{W1}. The $W$ boson mass, at tree level, is equal to $g\nu/2$, where $\nu=246\ \mathrm{GeV}$ is the vacuum expectation value of the Higgs field and $g$ denotes the weak isospin coupling parameter. If new particles are present, the $W$ boson mass is expected to receive additional loop corrections.

Moreover, The ATOMKI collaboration has reported a $\sim 7\sigma$ in the opening angle and invariant mass distributions of $e^+e^-$ pairs produced during the nuclear transition of the excited $^8Be^*$\cite{X1,X2}. These results have been confirmed also with other nuclei, like $^4He$.

It has been speculated that this anomaly can be interpreted as the emission of a protophobic gauge boson corresponding to a new $U(1)_X$ symmetry, the $X_{17}$ boson, with a mass of about $17 \ \mathrm{MeV}$ and which decays into $e^+e^-$ \cite{X3,X4}.
Several experiments and theoretical analyses have been performed to assess the viability of this new boson \cite{X5,X6,X7,X8,X9,X10,X10B,X11,X12,X13,X14}.
Based on the paper \cite{RS} we argue that the introduction of the new vector boson $X_{17}$ could, in principle, explain in a simple and elegant manner all the tensions we have cited above.
 
 \section{muon magnetic moment}
 
 \begin{figure}[t]
 \centering
\includegraphics[width=0.7 \textwidth]{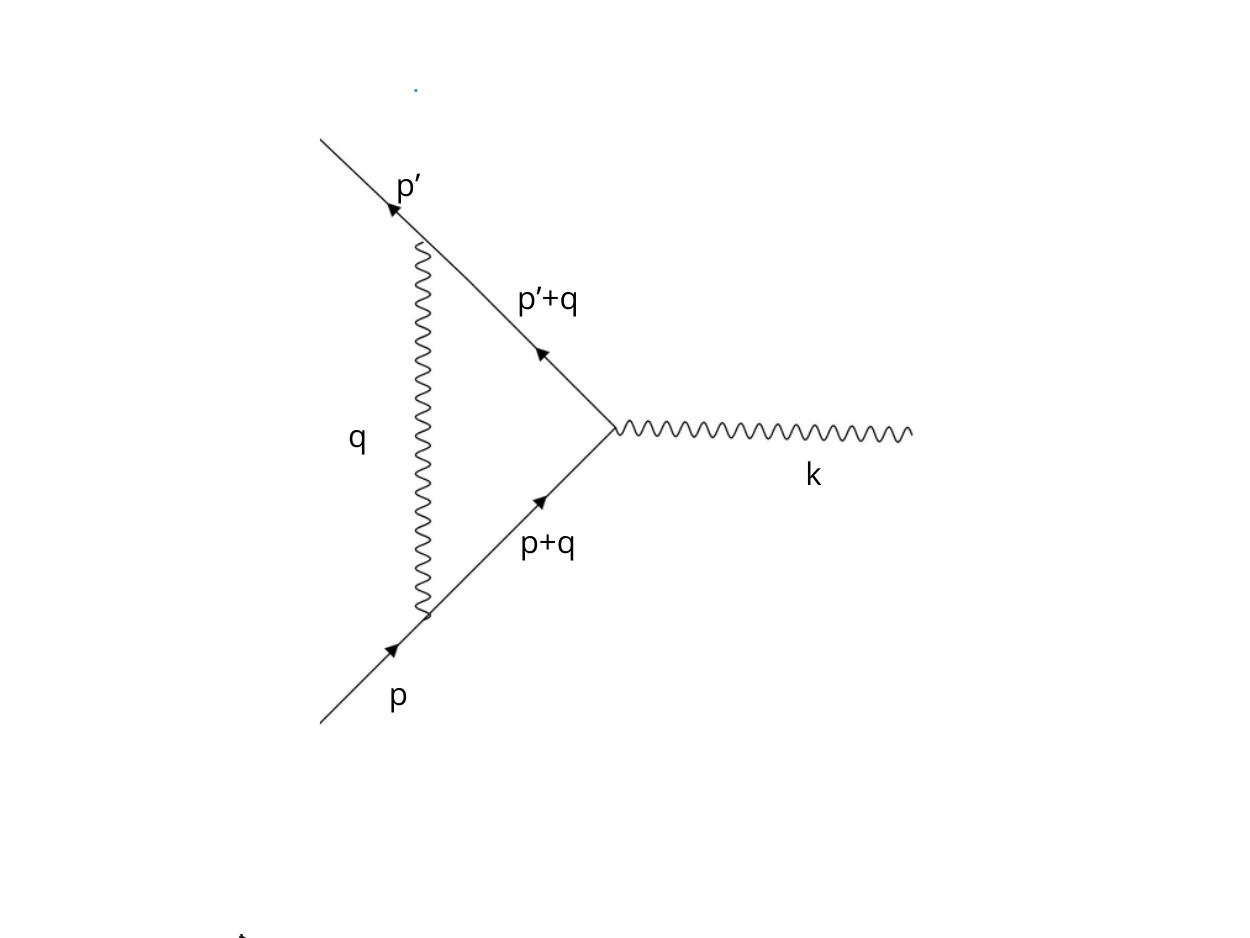}
\caption{Reference Feynman diagram for the calculation of the one loop correction to the magnetic dipole moment.}
\end{figure}
The presence of a new vector boson as $X_{17}$, which couples with the charged leptons, leads to a correction to the $g-$factor.
These correction emerge at one loop, so the relevant diagram is the one shown in Fig. 1.
The resulting correction to the $g-$factor is 
 \begin{equation}
 \label{deltaa}
    a_\mu^X= \frac{\alpha}{2\pi} \epsilon_\mu^2 \lambda_\mu \int_0^1 dx \frac{x^2(1-x)}{\lambda_\mu x^2-x+1}=\frac{\alpha}{2\pi} \epsilon_\mu^2 f(\lambda_\mu),
 \end{equation}
where $\epsilon_\mu$ is the coupling constant between $X_{17}$ and the muon and the adimensional parameter $\lambda_\mu=m^2_\mu/M^2_X$ is the ratio between the muon mass $m_\mu$ and the  $X_{17}$ mass $M_X$.
Since this correction depends on the ratio between the lepton mass and the boson mass, the introduction of $X_{17}$ could explain why the tension between the experimental results and theoretical predictions is greater for the muon.
When calculating the correction to the electron anomalous magnetic moment, the effect would be much smaller due to the smaller mass. Conversely, we can hypothesize that the correction for the tau lepton would be even larger.

Attributing the difference in the observed values of $a_\mu$ and the standard model computation to the $X_{17}$ correction, we now determine the upper bounds on $\epsilon_\mu$.
The relevant inequality we will use is
\begin{eqnarray}\label{momm}
 \delta a_\mu= a_{\mu , EXP} - a_{\mu, SM} \leq 2.51\times 10^{-9} \ ,
\end{eqnarray}
in which the latest available average values for the Standard Model prediction \cite{Patrignani2016} for $a_{\mu, SM}$ and the experimental result \cite{Abi2021} for $a_{\mu , EXP}$ are considered.
Notice that there are two possibilities: in principle the dark charges may be flavor-blind or flavor dependent.
However, it can be shown that \cite{Feng} flavor-blind couplings would lead to unobserved charged lepton oscillations.
Therefore we consider the case of flavor-dependent couplings.
 Considering a mass of the $X_{17}$ of $17\  \mathrm{MeV}$, the upper bound on the coupling between the $X_{17}$ and the muon, as resulting from the comparison between the inequality \eqref{momm} and Eq. \eqref{deltaa} is $|\epsilon_\mu| <2.154 \times 10^{-4}$. Notice that, since the correction of Eq. \eqref{deltaa} is quadratic in $\epsilon_\mu$, we cannot establish the absolute sign of $\epsilon_{\mu}$ from this analysis. 

\section{ Lamb shift}
We analyze the possible contribution of $X_{17}$ to the Lamb shift for muonic atoms.
The nonrelativistic potential between the proton and muon due to the $X_{17}$ exchange is:
\begin{equation}
\label{Pot}
    V_X(r)=\frac{\epsilon_\mu \epsilon_p}{e^2} \frac{\alpha e^{-M_X r}}{r},
\end{equation}
where $\epsilon_\mu$ can be estimated as shown above from the analysis of $g-2$ anomaly and $\epsilon_p$ is the coupling between the $X_{17}$ and the proton.
This potential gives an additional contribution to the Lamb shift in the $2S_{1/2}-2P_{3/2}$ transition, which, using first order perturbation theory, is given by:
\begin{equation}
\label{deltaH}
\begin{split}
    \delta E^H_X&=\int dr r^2V_X(r)(\|R_{20}(r)\|^2-\|R_{21}\|^2)\\
    &= \frac{\alpha}{2 a_H^3} \biggl(\frac{\epsilon_\mu \epsilon_p}{e^2}\biggr) \frac{f(a_H M_X)}{M_X^2} \ .
    \end{split}
\end{equation}
Here $f(x)=\frac{x^4}{(1+x)^4}$ and $a_{H}=(\alpha m_{\mu p})^{-1}$ is the Bohr radius of the system, $m_{\mu p}$ is the reduced mass of the muonic hydrogen and $R_{nl}$ are the radial wave function. Fixing the value of $\epsilon_{\mu}$ equal to the upper bound found from the previous analysis on the muon $g-2$ anomaly, and inserting the experimental value for $\delta E^{H}_{\mu}$, we can invert Eq. \eqref{deltaH} to deduce $\epsilon_p$. This yields an upper bound on the (modulus of the) coupling between the $X_{17}$ and the proton in terms of the $X_{17}$ mass $M_X$ and of the Lamb shift deviation $\delta E^{H}_{\mu}$, as depicted in Fig. 2. Since $\epsilon_p < 0$ for $\epsilon_{\mu} > 0$, the contour plot of Fig. 2 displays the lower bound on $\epsilon_p$, as a function of  the mass of $X_{17}$ and experimental data on muonic hydrogen  \cite{muon1, muon2}. The values are deduced by inverting Eq. \eqref{deltaH} in correspondence of $\epsilon_{\mu} = 2.154 \times 10^{-4}$, as derived from the analysis on the $g-2$ anomaly.  Considering a range of values for $M_{X}$ between $[16.7, 17.2] \ \mathrm{MeV}$, and for $\delta E_\mu^H$ between $[-0.363, -0.251]\ \mathrm{meV}$, the range of values for the lower bound on the coupling $\epsilon_p$ will be between $[-0.04260,-0.02982]$. Consequently the upper bound on $|\epsilon_p|$ ranges in $[0.02982,0.04260]$.

As remarked above, there is a sign ambiguity on $\epsilon_{\mu}$: the values of Fig. 2 are obtained by assuming a positive muon coupling $\epsilon_{\mu} > 0$, which leads to a negative proton coupling $\epsilon_{p} < 0$. The general conclusion is that $\epsilon_p$ and $\epsilon_{\mu}$ must have opposite sign in order to account for the Lamb shift anomaly in muonic hydrogen. As clear from Eq. \eqref{deltaH} and from Fig. 2, the lower bound on $\epsilon_p$ is linear in the Lamb shift deviation $\delta E_\mu^H$ and approximately quadratic in the boson mass $M_X$. 
In an analogous way, it is possible to derive an upper bound on $\epsilon_n$, the coupling of the $X_{17}$ to the neutron.
In Fig. 3, we report a contour plot of the upper (lower) bound on the coupling constant $\epsilon_n$, depending on its sign $\epsilon_n > 0$ ($\epsilon_n < 0$) as a function of experimental data on muonic hydrogen \cite{muon1, muon2} and muonic deuterium \cite{muon3, muon4} and in correspondence with $M_X = 17 \ \mathrm{MeV}$. Considering a range of values for for $\delta E_\mu^H$ between $[-0.363, -0.251]\ \mathrm{meV}$, and for $\delta E_\mu^D$ between $[-0.475, -0.337]\ \mathrm{meV}$, the range of values for the coupling $\epsilon_n$ will be between $[-0.010, 0.010]$. Overall $|\epsilon_n| \leq 10^{-2}$.
\begin{figure}[t]
\centering
\includegraphics[width= 0.6 \textwidth]{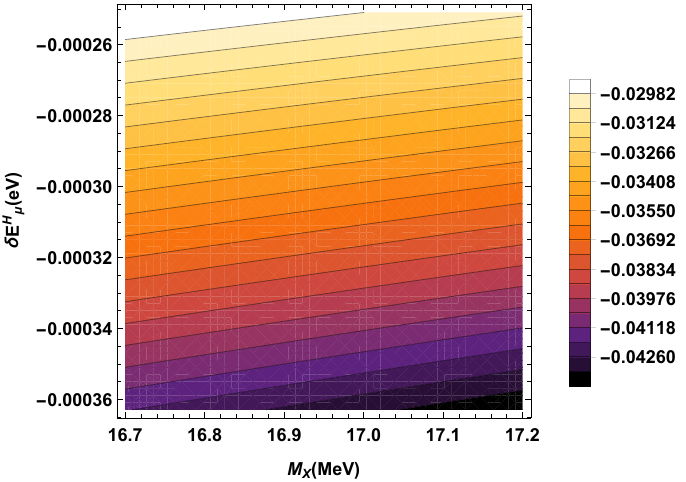}
\caption{ (Color online) Contour plot of the lower bound for the coupling constant $\epsilon_p$ between the $X_{17}$ and the proton as a function of the mass of $X_{17}$ and the experimental data on muonic hydrogen Lamb shift  $\delta E_\mu^H$ \cite{muon1, muon2}, as obtained from equation \eqref{deltaH}. The legend on the side shows the values of the coupling constant $\epsilon_p$ for different values of $M_X$ and $\delta E_\mu^H$. We consider $\epsilon_{\mu} \simeq 2.154 \times 10^{-4}$ .}. 
\end{figure}

\begin{figure}[t]
\centering
\includegraphics[width= 0.6 \textwidth]{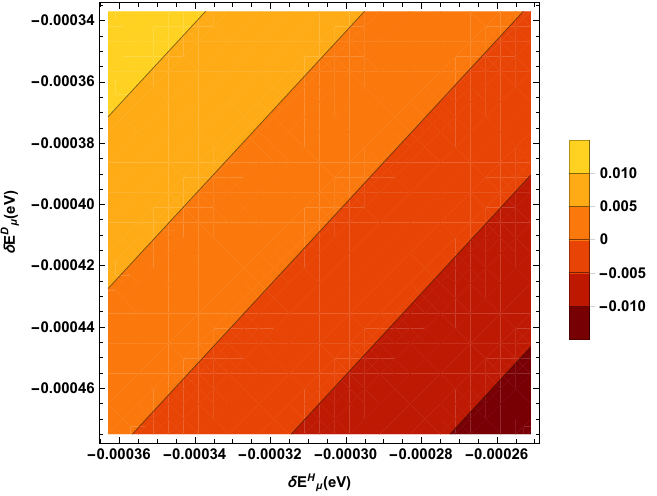}
\caption{ (Color online) Contour plot of the upper (lower) bound for the coupling constant $\epsilon_n > 0$ ($\epsilon_n < 0$)between the $X_{17}$ and the neutron in terms  experimental data on  muonic hydrogen shift $\delta E_\mu^H$  \cite{muon1, muon2} and muonic deuterium $\delta E_\mu^D$ \cite{muon3, muon4}. The legend on the side shows the values of the coupling $\epsilon_n$ for different values of $\delta E_\mu^H$ and $\delta E_\mu^D$. We have used $|\epsilon_{\mu}| \simeq 2.154 \times 10^{-4}$ and we have assumed $\epsilon_\mu>0$. The coupling to the proton $\epsilon_p$ is fixed according to Eq. \eqref{deltaH}, ranging in the values shown in Fig. 2. The boson mass is here fixed to the value $M_X = 17 \ \mathrm{MeV}$.}
\end{figure}

\section{$W$ mass}
For this analysis we follow the method used in \cite{W00} for the kinetic mixing between the hypercharge boson $B$ and the dark photon, suitably modified for the $X_{17}$.
The shift on the $W$ mass is given by:
\begin{equation}\label{Wmassshift}
    \Delta M_W=M_W-M_{W,SM}=-\frac{M_{W,SM}s_W^2\xi^2}{2(c_W^2-s_W^2)^2(1-r^2)},
\end{equation}
 where $c_W=\cos{\theta_W}$,$s_W=\cos{\theta_W}$, $\theta_W$ is the Weinberg angle, $\xi$ is the kinetic mixing parameter and  $r=\frac{M_Z}{M_X}$.
 By setting $\Delta M_W\leq 10 \ \mathrm{MeV}$, corresponding to the uncertainty from the latest measurements \cite{W1}, we derive, from Eq. \eqref{Wmassshift}, an upper bound on the kinetic mixing parameter: $|\xi|<2.2 \times 10^{-2}$.
 
\section{Conclusions}
We have investigated the effect of $X_{17}$ on $g-2$ muon anomaly and Lamb shift.
We calculated the upper bounds on the couplings of this boson with the muon, proton, and neutron. Our results revealed an ambiguity in the signs of the coupling costants, suggesting that the muon, proton, and neutron carry opposite charges under a hypothetical $U(1)_{X}$ symmetry. 
Analyzing the kinetic mixing between the $X_{17}$ boson and the $U(1)_Y$ gauge boson, we established an upper limit on the kinetic mixing constant, constrained by the most recent experimental uncertainties.

\section*{Acknowledgements}
Partial financial support from MIUR and INFN is acknowledged.
A.C. also acknowledges the COST Action CA1511 Cosmology
and Astrophysics Network for Theoretical Advances and Training Actions (CANTATA).

\end{document}